\documentclass[aps,twocolumn,showpacs]{revtex4}

\usepackage{amsmath}
\usepackage{graphicx}
\def\lsim{\lower.8ex\hbox{$\buildrel<\over\sim$}}
\def\gsim{\lower.8ex\hbox{$\buildrel>\over\sim$}}

\begin{document}

\date{January 2nd, 2004}

\title {Colloidal brazil nut effect in sediments of binary charged
    suspensions}

\author  {Ansgar Esztermann, Hartmut L{\"o}wen }
\affiliation  {Institut f{\"u}r Theoretische Physik II,
Heinrich-Heine-Universit{\"a}t D{\"u}sseldorf, Universit{\"a}tsstra\ss e 1, 
D-40225 D{\"u}sseldorf, Germany}

\begin{abstract}
Equilibrium sedimentation density
profiles of charged binary colloidal suspensions
are calculated by computer simulations and density functional theory.
For deionized samples, we predict a colloidal ``brazil nut'' effect: 
heavy colloidal particles sediment on top
of the lighter ones provided that their  mass per  charge
 is smaller than that of the lighter ones.
This effect  is  verifiable in settling experiments.
\end{abstract}


\maketitle

Binary systems of granular matter separate upon shaking in gravity,
so that the larger particles lie on top of the smaller ones even if they
are heavier and denser than the latter. This is due to a sifting mechanism
in which tiny grains filter through the interstices between the large particles
which is well-known as ``brazil nut'' effect:
in a jar of mixed nuts or in a package of cereal, 
the largest species rises to the top \cite{Williams,brazil0}.
This clearly distinguishes granular matter from ordinary fluids where
the rising species is controlled by Archimedes' law. Understanding the full
details of the brazil nut effect is still a problem; recently even a reverse brazil nut effect of
large light grains sinking in a granular bed has 
been predicted \cite{brazil1,brazil2,brazil3}
and verified in experiments \cite{brazil4}. 

In this letter, we report on equilibrium density profiles of binary charged colloidal fluids
(``macroions'') under gravity. Using extensive
Monte-Carlo computer simulations of the ``primitive'' model \cite{Linse}
of strongly asymmetric electrolytes  and a  density functional theory,
we predict that the heavier particles sediment on top of the lighter ones 
provided the charge per mass of the heavier particles is higher. 
In analogy to granular matter, we call this counter-intuitive phenomenon a
{\it colloidal brazil nut effect}.
It is generated by the entropy of the microscopic
counterions in the solution, which are coupled to the macroions by strong
Coulomb binding.
Clearly, though this effect is  qualitatively similar to
the granular brazil nut effect insofar as heavy particles are on 
top of lighter ones,
its  physical origin is different:
first, the particle charge (and not the size) is crucial.
Second, the colloidal brazil nut effect
is a pure  equilibrium phenomenon while the granular brazil nut effect happens 
intrinsically in non-equilibrium.
The colloidal brazil nut effect can be verified, e.g., in depolarized-light 
scattering or real-space experiments on 
sediments of strongly deionized binary charged suspensions 
\cite{Piazza,Chaikin1}. Similar techniques have been used
to measure one-component colloidal density profiles \cite{Piazza,Philipse}
where deviations from the ideal barometric law \cite{Einstein}
are still an ongoing debate \cite{Biben,Simonin,Loewen_JPCM,Tellez_Biben,bib:vanRoij}.

We simulate the asymmetric ``primitive'' model of binary charged suspensions 
 in which
the solvent only enters via a continuous dielectric background with 
permittivity $\epsilon$ but all charged particles
(two species of negatively charged macroions and microscopic counter- and coions)
are treated explicitly at constant temperature $T$. With $Z_1e$, $Z_2e$, $-qe$ and $\sigma_1$,
$\sigma_2$, $\sigma_{c}$ denoting the charges and the diameters 
of the two colloidal species and the microions, the interaction between the charged particles
is given as a combination of Coulomb forces and excluded volume of the hard particle cores.
Here, $e$ is the electron charge.
For simplicity we assume that the co- and counterions of the salt solution have the same
valency and the same hard core diameter $\sigma_c$.
We consider a finite system of $N_1$, $N_2$ charged macroions and corresponding number of counterions
(fixed by global charge neutrality) 
plus  salt ions of bulk concentration $c_s$. The simulation box 
is rectangular with lengths $L_x=L_y$ and $L_z=32 \,L_x$ 
in the three different spatial directions with periodic boundary
conditions in $x$ and $y$ direction and finite length in $z$-direction.
Hard walls are placed at $z=0,L_z$ and gravity with acceleration $g$ points
along the $-z$ direction. Only the colloidal particles with their
buoyant masses $m_1$ and $m_2$ are subject to  gravity,
whereas the microions are not.

A Monte-Carlo simulation is performed in the canonical ensemble 
with the long-ranged Coulomb interaction 
treated via Lekner sums \cite{bib:lekner}.
Typically $10^3$ Monte Carlo moves per particle were performed 
for equilibration and it took an additional  $10^4$ Monte Carlo moves per particle
to gather statistics. Finite system size effects were carefully checked
by changing all lateral linear dimensions by a factor of 4. This
means that we have changed the total number of colloidal particles $N_{1}+N_{2}$
in the range of 12--200.
We have calculated the inhomogeneous 
$z$-dependent averaged density profiles $\rho_1 (z)$,
$\rho_2 (z)$, $\rho_3 (z)$ and $\rho_4(z)$ of the two macroions and the counter- and coions.
Data are shown for the largest system size where $N_1=N_2=100$.

Besides the ratios $Z_{1}/q$, $Z_{2}/q$, $m_{1}/m_{2}$, $\sigma_{1}/\sigma_{2}$, $\sigma_{c}/\sigma_{2}$,
the system is characterized by two partial area densities 
$n_i = N_i/L_x L_y$ ($i=1,2$), the Bjerrum length $\lambda_B= q^{2}e^{2}/\epsilon kT$
($kT$ denoting the thermal energy), the gravitational length
$\ell_{2}= kT/m_{2}g$ of the second particle species, and the salt concentration.
In order to reduce the parameter space, we have assumed throughout the simulations
monovalent microions $(q=1)$, and the same hard core diameter of the macroions
$\sigma =\sigma_1 =\sigma_2$ which serves as a natural length scale. We further fixed
$Z_{2}=15$, $\lambda_B=\sigma_c=\sigma/128$ and 
$n_1=n_{2}=0.1/\sigma^{2}$. This corresponds to typical parameters for
low-charge aqueous suspensions.
We varied the colloidal
charge $Z_{1}$, the colloidal mass ratio $m_{1}/m_{2}$ (with $m_1>m_{2}$), the gravitational
length $\ell_{2}$, and the salt concentration.

\begin{figure}
  \includegraphics[width=8.5cm]{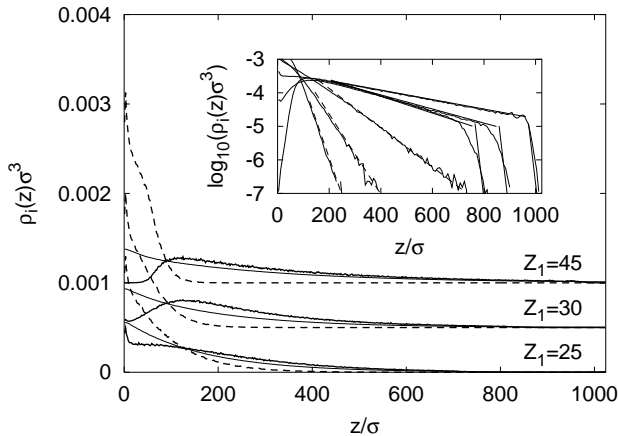}
  \caption{Density profiles of macro- and counterions for $Z_1=45,30,25$ 
  (solid lines; top to bottom). The second colloidal component
  is shown as a dashed line.
  Counterion densities (thin lines) have been divided by $Z_1+Z_2$.
  For clarity, curves pertaining to different
  simulation runs have been shifted by $5\times10^{-4}$ with respect
  to each other. The parameters are: $m_1/m_2=1.5$, $\ell_{2}/\sigma=10$, $c_{s}=0$.
  Inset: Semi-logarithmic plots of the same colloidal density profiles
 with the slopes predicted from theory.
  }
  \label{fig:profile}
\end{figure}

In the salt-free case,
density profiles for the two macroions and the counterions
are shown in Figure~\ref{fig:profile} for $m_1/m_2=1.5$ and three different
macroion charges $Z_1=45,30,25$. 
For large heights $z$, the heavy particles
(solid curve) are on top of the lighter ones (dashed curve). This
colloidal brazil nut effect is getting  stronger for
increasing charge asymmetry between the colloidal particles: 
for the highest charge $Z_{1}=45$,
the two macroion species are almost completely separated.
To quantify the brazil nut effect, we define
-- in analogy to the ordinary 
brazil nut problem \cite{brazil3} --  a mean (or sedimentation)
height of the two profiles via
\begin{equation}\label{eq:mean_h}
h_i = \frac{\int_{0}^{\infty} z \rho_i (z) \,dz}{  \int_{0}^{\infty} \rho_i(z)\,dz }
  \qquad i=1,2
\end{equation}
By definition, the colloidal brazil nut effect occurs
if $h_1>h_2$, but there is no brazil nut effect (or
a ``reverse'' brazil nut effect) in the opposite case, $h_1\leq h_2$.

Let us now describe the properties of the density profiles for increasing $z$
qualitatively and put forward a simple theory to describe the basic features.
Close to the hard container bottom at $z=0$, small correlation effects are visible
as a density shoulder of the low-charge particles while
high charge particles are depleted from the wall.
The reason is a combination of  a pure interface effect and Archimedes'
principle. First, we have checked by
simulation that depletion of high-charge particle
persists for zero gravity. Second, by crudely mapping the interacting binary colloidal mixture onto
one with effective hard interaction cores, the 
high-charge particles will have a larger diameter and hence Archimedes'
principle will lift the high-charge particles in the sea of small ones provided
their mass density is smaller \cite{Walliser}. For our parameters, however, the latter
effect is small and confined to regions close to the wall.
More importantly, as revealed by the semi-logarithmic plot in the inset
of Figure~\ref{fig:profile},
there is an exponential decay of the two colloidal profiles
for intermediate heights associated with two different decay lengths for the two colloidal species.
The decay length for the heavier but high-charge particles is larger than that for
the low-charge particles. This gives the most significant contribution in the integral
(1) of the mean height.
Finally, for large $z$, there is cross-over towards another exponential decay
involving the gravitational length $\ell_{1} = kT/m_{1}g$
for the high-charge particles. It is important to note that the density profiles
fulfill {\it local charge neutrality} throughout the whole sample 
except at the container bottom ($z=0$) and at very large heights $z$. But even
there the excess charge separated is small \cite{footnote}.

Our  theoretical explanation for the colloidal
brazil nut effect is based on a simple density functional approach.
The free energy ${\cal F}$ per unit area,
which is a functional of the inhomogeneous density
fields $\rho_1 (z), \rho_2 (z), \rho_3 (z)$
of the macro- and counterions,
splits into the gravitational energy, the entropy of the three species
and all Coulomb contributions. The latter are approximated within
a mean-field-type  Poisson-Boltzmann theory 
\cite{Biben,Simonin,Loewen_JPCM,Tellez_Biben,bib:vanRoij}. Hence:
\begin{equation}
\begin{split}\label{eq:functional}
{\cal F}  [&\rho_1 (z), \rho_2 (z), \rho_3 (z)]  =
  \sum_{\nu =1}^2 \int_0^{\infty}\!\! dz\, m_{\nu} g z  \rho_{\nu} (z)\\
  &+\sum_{\nu =1}^3  \int_0^{\infty} dz\,  
    kT \rho_{\nu } (z) ( \ln ( \Lambda_{\nu }^3 \rho_{\nu } (z)) -1 ) \\
  &+\frac{1}{ 2} \int d^2r' \int_0^{\infty}\!\! dz\int_0^{\infty}\!\! dz'\,
    \frac{{\rho_{t} (z)\rho_{t} (z')}}{\epsilon \sqrt{ r'^2 + (z-z')^2}} 
\end{split}
\end{equation}
Here,  $\Lambda_{\nu }, (\nu=1,2,3)$, are Lagrange multipliers
which ensure that the overall colloidal densities
per unit area equal the prescribed number densities, 
i.e.  $\int_0^{\infty} dz \  \rho_i (z) = n_i$ ($i=1,2$).
$\Lambda_3$ is fixed by {\it global} charge neutrality,
and 
$\rho_t(z)=  Z_1e \rho_1(z)+  Z_2e \rho_2(z) -  qe  \rho_3(z)  $
is the total {\it local} charge density of the system.
The functional ${\cal F}$ is minimal for the physically realized 
equilibrium density 
profiles. 

We discuss two different cases of weak and strong Coulomb coupling
subsequently. For weak Coulomb coupling, which is realized
for very large heights $z$ where the densities are extremely small,
 one may neglect the third term
on the right hand side of Eq. (2).
Then, the minimization of
${\cal F}$ yields colloidal density profiles which follow
the traditional barometric law
$\rho_{i}(z) \propto \exp ( -z/\ell_i)$ $(i=1,2)$.
Strong Coulomb coupling, on the other hand, will impose
{\it local} charge neutrality $\rho_t(z)=0$ which implies
that the counterion density field is enslaved to that of the macroions.
Minimizing ${\cal F}$ in this limit with respect to the two
colloidal densities only, again yields an
exponential decay
$\rho_{i}(z) \propto \exp ( -\gamma_{i} z)$ $(i=1,2)$
but with  inverse decay lengths $\gamma_{i}$ which are smaller
than $1/\ell_i$. These decay lengths $\gamma_{i}$
turn out to be as follows:
Let $\alpha$ be the index $i$ for which the mass per charge
ratio $\frac{m_{i}}{Z_{i}/q +1}$ is {\it minimal}
and let $\beta$ be the index complementary to $\alpha$, i.e.\ $\beta=1$ if $\alpha =2$
and $\beta=2$ if $\alpha =1$. Then
\begin{equation}\label{eq:top-gamma}
\gamma_{\alpha} = {\frac{m_{\alpha}}{  Z_{\alpha} / q +1}}\frac{g}{kT}   
\end{equation}
and
\begin{equation}\label{eq:bot-gamma}
\gamma_{\beta} = \left(m_\beta-m_\alpha\frac{Z_\beta/q}{Z_\alpha/q+1}\right)
    \frac{g}{kT} >  \gamma_{\alpha}  
\end{equation}
with $  \gamma_{\beta}  \ge  {\frac{m_{\beta}g}{kT ( Z_{\beta} / q +1)}} \ge \gamma_{\alpha}$.
This second case of strong Coulomb coupling 
will be realized for heights $z$ where correlations between the ions
are small (which justifies the mean-field approximation) but where local charge neutrality is still
valid. The physical reason for the much slower decay of the colloidal density profiles
in the second
regime results from the counterion entropy which tends to delocalize
the counterions. However, since the macroions are coupled to the counterions
 due to the constraint of local charge neutrality, they are lifted upwards
together with their counterions.
Assuming that the main contribution in the integrand of the right-hand-side of (1)
comes from the second regime,
the mean heights are given by $h_{\alpha} = 1/\gamma_\alpha$ and $h_{\beta} = 1/\gamma_\beta$.
A  brazil nut effect
occurs when $\alpha=1$ and $\beta=2$. Hence the transition towards the brazil nut effect
happens at $\frac{m_{1}}{Z_{1}/q +1} =\frac{m_{2}}{Z_{2}/q +1}$.

\begin{figure}
  \includegraphics[width=8.5cm]{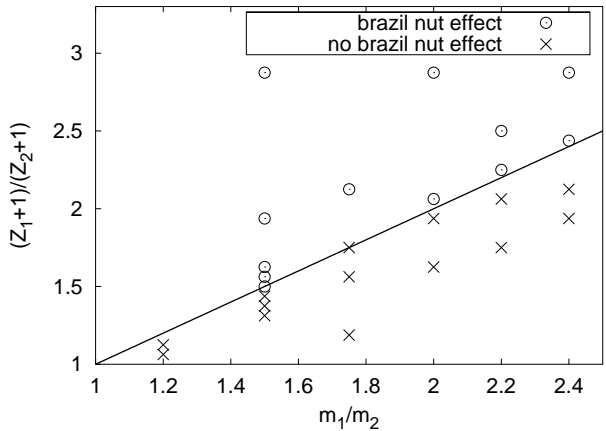}
  \caption{Crossover towards the brazil nut effect. Shown are different
    charge asymmetries $(Z_1+1)/(Z_2+1)$ and mass ratios $m_1/m_2$ for
    which the brazil nut effect was (circles) or was not observed
    (crosses). The straight line shows the theoretical prediction.
    The parameter combinations were obtained by changing
    $\ell_{2}$ and $Z_{1}$ is the range of $4\dots20\, \sigma$
    and $16\dots45$, respectively.}
  \label{fig:transition}
\end{figure}

Let us now test the prediction of the theory
against our simulation data. First, the slopes in the inset
of Figure~\ref{fig:profile} confirm the inverse decay lengths $\gamma_{\alpha}$
and $\gamma_{\beta}$ perfectly for a large range of intermediate heights; the theoretical predictions 
for the slopes as given by Eqs.\ (3) and (4) are shown as thick lines. 
The crossover to the bare gravitational length
for large $z$ is confirmed for the high-charge particles and was
found to be  pretty sharp.
Consistently with the theoretical assumption, local
charge neutrality in the intermediate regime is fulfilled.
Second, we have tested the location of the transition
towards the colloidal brazil nut effect by systematically
varying the  mass and charge ratio.
The results are summarized in Figure~\ref{fig:transition}.
The theory, which predicts the transition  at $(Z_1+1)/(Z_2 +1) = m_1/m_2$, 
is shown as a straight line there.
All parameter combinations simulated are indeed separated by this theoretical
prediction, confirming our simple theory. This is remarkable as any
wall or bulk correlation effects are neglected in the theory.
\begin{figure}
  \includegraphics[width=8.5cm]{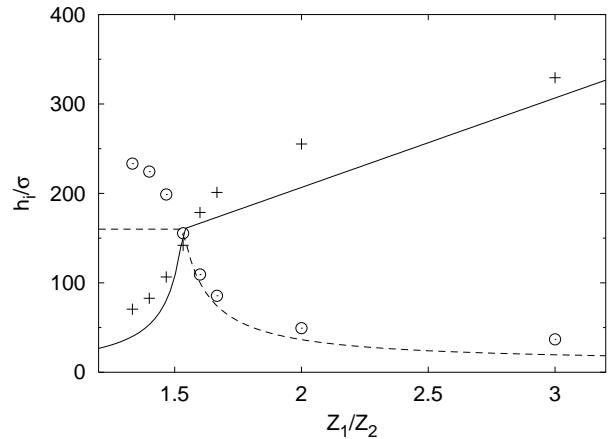}
  \caption{Mean heights $h_{i}$ of heavy ($+$) and light ($\odot$)
    particles as a function of  $Z_1$ for the parameters of
    Fig.~\ref{fig:profile}.
    The lines show the theoretical predictions.}
  \label{fig:height}
\end{figure}

In order to elucidate this further, we have compared
the theoretical predictions $1/\gamma_1$, $1/\gamma_2$ for the heights 
with
the simulation data in a situation where
the transition line was crossed. In Figure~\ref{fig:height}
the heights are shown as a function of a varied charge $Z_{1}$. The 
simulation data
reveal that the variation of both heights $h_i$ with $Z_1$ is 
large close to the transition and becomes maximal at the transition.
In the theory, this feature is reproduced and accompanied by 
 a generic cusp  at the transition. Though
the cusp is smeared out and  in general the heights
are larger  in the simulation data 
due to the density reduction close to the wall, there is still
semi-quantitative agreement. The rapid variation of the heights
at the transition  implies that the location of the transition
towards the brazil nut effect is very robust 
explaining the validity of the theory in Figure~\ref{fig:transition}.

\begin{figure}
  \includegraphics[width=8.5cm]{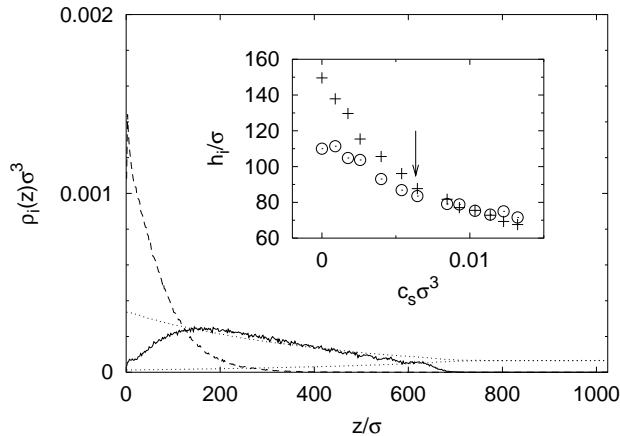}
  \caption{Density profiles for  added salt. 
    The counter- and coion densities have been
    divided by $Z_1+Z_2$, the coion density is smaller than that of the counterions. 
The parameters are
    $Z_1=45$,  $m_1/m_2=2$, $\ell_{2}=20\,\sigma$,
    and $c_s\sigma^{3}=2.5\times10^{-3}$.
    Inset:
    Sedimentation  heights $h_{i}$ of heavy ($+$) and light ($\odot$)
    particles as a function of reduced salt concentration for
    $Z_1=32$, 
    $m_1/m_2=2$, and $\ell_{2}=5\,\sigma$.
    The theoretical estimate of the threshold salt concentration is
    marked by an arrow.
  }
  \label{fig:salt}
\end{figure}

Finally we address the case of added monovalent salt.
Density profiles for the two macroion species
and the counter- and coions are presented in Figure~\ref{fig:salt}.
Addition of salt reduces the brazil nut effect; in the inset of
Figure~\ref{fig:salt} the heights are plotted versus added salt concentration.
The threshold salt concentration $c_s^{(0)}$ 
needed to reverse the brazil nut effect is estimated as
$c_s^{(0)} \approx Z_{1}\rho_{1}(z_{0}) +Z_{2}\rho_{2}(z_{0})$
which is the counterion concentration at the position $z_0$ 
where the two colloidal density profiles
cross in the salt-free case. This salt concentration is indicated as an arrow in the inset
of Figure~\ref{fig:salt}, confirming the validity of the estimate.
Though the brazil nut effect is proven to remain stable with respect to added salt,
 it will only show up for deionized solutions.
Highly charged suspensions in 
non-polar solvents with a small dielectric constant $\epsilon$
and low impurity ion concentrations  \cite{Royall} are
promising candidates to exhibit a strong brazil nut effect.

In conclusion, we predict an analog of the granular brazil nut effect in 
 equilibrium
sediments of charged suspensions,
which is generated by an entropic charge lifting 
due to the Coulomb coupling to the counterions. 
An experimental verification of the levitation  should be possible employing
depolarized light scattering
\cite{Piazza} or confocal microscopy \cite{Royall}.
The simulated charge asymmetries correspond to micelles and proteins rather than highly
charged colloids but the simple theory which was confirmed by the simulations is
applicable  for arbitrary  charges. 

The lifting effect is immediately generalizable to sediments of 
solutions which are polydisperse in mass and charge \cite{Xu}.
Therefore the  colloidal brazil nut effect has important biochemical implications
for the  separation of polydisperse biological matter,
such as protein solutions. Analytical sedimentation
is typically used in an ultracentrifuge
to separate different species \cite{centrifuge}.
The colloidal brazil nut effect implies that the separation is sensitive
to the mass per charge but not to the mass itself.

We are grateful to T.~Palberg, F.~Scheffold, A.~Philipse, C.N.~Likos, R.~van~Roij,
  and R.~Piazza for helpful discussions.
Financial support within the
Deutsche Forschungsgemeinschaft (wetting priority program and SFB TR6)
is gratefully acknowledged.


\begin{thebibliography}{1cm}
\bibitem{footnote} The charge non-neutrality at the system
boundaries will result in a small electric field as recently predicted by
theory in Ref.~\cite{bib:vanRoij} and also confirmed by 
simulation of the one-component colloidal system
(A.-P.~Hynninen, R.~van~Roij, M.~Dijkstra, private
communication). The presence of this field, however, will change
neither the density field for intermediate $z$
nor the general conclusions of our analysis which assumes local charge neutrality.
\end{thebibliography}
\end{document}